# Influence of electron cooling on the polarization lifetime of a horizontally polarized storage ring beam

S. Karanth,[1] E. Stephenson,[2] A. Wrońska,[1] G. Ciullo,[3] S. Dymov,[3,4] R. Gebel,[5]
G. Guidoboni,[3] V. Hejny,[5] A. Kacharava,[5] I. Keshelashvili,[5] P. Lenisa,[3] A. Lehrach,[5,6]
B. Lorentz,[5] D. Mchedlishvili,[7] A. Nass,[5] N. Nikolaev,[8,9] A. Pesce,[5] J. Pretz,[5,6,10] D. Prasuhn,[5]
F. Rathmann,[5] A. Saleev,[5,11] Y. Senichev,[5] V. Shmakova,[4,5] H. Ströher,[5,6] R. Talman,[12] Yu. Valdau,[5]
C. Weidemann,[5] and P. Wüstner[13]

(for the JEDI Collaboration)

[1]*Marian Smoluchowski Institute of Physics, Jagiellonian University, 30348 Cracow, Poland*
[2]*Indiana University Center for Spacetime Symmetries, Department of Physics, Bloomington, Indiana 47405, USA*
[3]*Physics and Earth Sciences Department of the University of Ferrara and INFN of Ferrara, 44122 Ferrara, Italy*
[4]*Laboratory of Nuclear Problems, Joint Institute for Nuclear Research, 141980 Dubna, Russia*
[5]*Institut für Kernphysik, Forschungszentrum Jülich, 52425 Jülich, Germany*
[6]*JARA-FAME (Forces and Matter Experiments), Forschungszentrum Jülich and RWTH Aachen University, 52056 Aachen, Germany*
[7]*High Energy Physics Institute, Tbilisi State University, 0186 Tbilisi, Georgia*
[8]*L. D. Landau Institute for Theoretical Physics, 142432 Chernogolovka, Russia*
[9]*Moscow Institute for Physics and Technology, 141700 Dolgoprudny, Russia*
[10]*III. Physikalisches Institut B, RWTH Aachen University, 52056 Aachen, Germany*
[11]*Samara National Research University, 443086 Samara, Russia*
[12]*Cornell University, Ithaca, New York 14850, USA*
[13]*Zentralinstitut für Engineering, Elektronik und Analytik, Forschungszentrum Jülich, 52425 Jülich, Germany*

**Abstract**

A previous publication has shown that the in-plane polarization (IPP) component of a polarized 0.97-GeV/c deuteron beam in the COSY storage ring may acquire a polarization half-life in excess of 1000 s through a combination of beam bunching, electron cooling (prior to any spin manipulation), sextupole field adjustment, and a limitation of the beam intensity. This paper documents further tests pointing to additional gains in the IPP lifetime if cooling is active throughout the beam store.

## I. Introduction

A series of papers [1-5] from the cooler synchrotron COSY [6] show progress toward using a polarized beam in a storage ring to search for an electric dipole moment (EDM) on the charged particles in the beam. The signature of the EDM is the rotation of the beam's polarization, which is aligned with the EDM, about the radial direction in response to the torque produced by the particle-frame radial electric field [7]. The search begins with the polarization aligned along the direction of the beam velocity. A positive EDM signal is the slow growth of a vertical (out of the ring plane) component of the beam polarization. Rotation of the polarization relative to the velocity in the ring plane is suppressed by choosing the correct combination of the electric and magnetic fields that produce bending in the storage ring arcs. The tendency of the in-plane polarization (IPP) components to disperse limits the time available to observe an EDM. Thus, a long IPP lifetime is one of the requirements for a sensitive storage ring search. Guidoboni *et al.* [3] reports that the polarization lifetime in the horizontal plane may be extended through a combination of beam bunching, electron cooling (prior to rotation of the polarization into the ring plane), sextupole corrections to the magnetic fields of the ring, and a limitation on beam current. A long IPP lifetime is also required for studies [8] at COSY that make use of an RF Wien filter to obtain sensitivity to the deuteron EDM by measuring the orientation of the beam's invariant spin axis.

The reported IPP half-life [3] of 1173 ± 172 s characterizes the measurements shown in Fig. 1 (reproduced from Ref. [3], Fig. 4) for a deuteron beam of 0.97 GeV/c. In COSY, this momentum corresponds to a beam revolution frequency of 750 kHz.

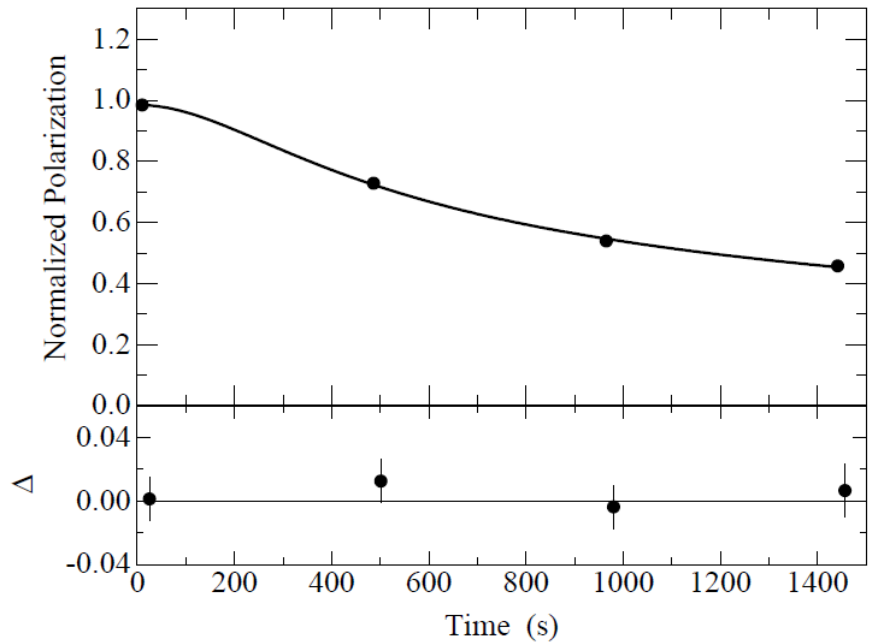

Figure 1: Measurements of the normalized horizontal polarization as a function of time in the beam store along with a fitted curve used to estimate the polarization lifetime. The data is normalized to the vertical polarization of the beam before it was rotated into the ring plane. The curve represents the polarization of a Gaussian distribution of spin tunes as it evolves in time. The lower panel shows the residuals of the fit. (Taken from Ref. [3].)

The spin tune of the beam, $v_S = G\gamma$ = -0.161 (G is the deuteron magnetic anomaly), means that the polarization rotates about one radian for every turn around the storage ring. So bunching is essential so that all particles make each orbit together without any losing or gaining a turn. The polarization, which is injected in the vertical direction, is rotated into the horizontal plane using an RF solenoid operating on the $(1 - G\gamma)$ harmonic of the revolution frequency for a brief period of time. Before that happens, the beam is electron cooled so that small variations in γ that would depolarize the beam are reduced to a minimum. As data acquisition begins, a small amount of white noise is imposed on the beam so that some particles are driven to a block carbon target next to the beam where they scatter into the polarimeter detectors, a process that uses up the beam. In order to preserve the beam for the time needed to characterize the IPP lifetime, the beam was sampled by the polarimeter only during four short time intervals in the beam store. The polarizations are represented by the four data points.

Figure 1 also contains a curve that was used to interpolate among the data points and obtain estimates of the polarization lifetime. The curve represents the polarization of a Gaussian distribution of spin tunes as it evolves in time. The lifetime is quoted as the time required for the normalized polarization to fall to a pre-determined value assuming that at t = 0 s the polarization is complete (equal to one). In addition to the half-life, Ref. [3] also quoted the time for the polarization to fall to 60.6% (Gaussian width, 782 ± 117 s) and 1/e (exponential width, 2280 ± 336 s) of its initial value of one. Each of these values were obtained from the curve. Errors were estimated from the process of fitting the curve to the measurements.

The IPP lifetime is a property of the distribution of spin tunes in the beam. The spin tune varies from one particle to another due to small differences in the relativistic factor γ. These variations in the bunched beam are associated with transverse oscillations, both vertically and horizontally, about the reference orbit. (Spin tune spreading caused by the variation of gamma accompanying particle momentum spreading cancels over each synchrotron oscillation period.) Each oscillation creates a longer path for the particle around the machine. When bunching maintains the revolution frequency for all particles, those on longer paths travel faster, thus increasing γ. Over time, the spin direction in the ring plane for particles with different speeds will diverge, leading to a loss of spin coherence, or beam polarization.

Electron cooling reduces the size of the beam in phase space. An example of these effects on the beam were illustrated in Ref. [2]. This also reduces the amplitude of betatron oscillations and the spread of spin tunes, thus extending the IPP lifetime. In the run shown in Fig. 1, electron cooling was on between the time that the stored beam reached full energy (about 2 s) and a point 75 s after beam injection. Afterward, the beam phase space slowly expands due to interactions with residual gas and, as described below, extraction of the beam onto the polarimeter carbon block target. At about 80 seconds, the polarization was rotated from the vertical direction into the ring plane through the use of an RF solenoid running for about 2 seconds on the $1 - G\gamma$ harmonic (871 kHz) [9].

Measurements of the beam polarization involved exciting vertical betatron oscillations in the beam by applying a white noise electric field to vertical strip line plates [1]. This brought beam deuterons to a carbon block that was inserted 3 mm above the beam reference orbit and, through scattering, to an array of polarimeter detectors. Electron cooling, which competes with this heating process, was not operating during the IPP measurement time.

In parallel with data acquisition for the polarization lifetime measurement of Fig. 1, sextupoles located in the arcs of the COSY ring were altered. Adding sextupole fields to the quadrupole fields already used for beam focusing creates a change of the reference orbit, or center position of the transverse (betatron) oscillation, that is proportional to the squared amplitude of the oscillation. This arises in the horizontal plane because the sum of quadrupole and sextupole fields creates a restoring force on either side of the beam that is not symmetric about the beam. This shifts the reference orbit to a new equilibrium point. If done correctly in the arcs of the ring, this change can lead to a shortening of the orbit length that compensates for the lengthening due to the oscillation itself. This change reduces the difference of spin tunes among the particles, leading to particularly long IPP lifetimes [3,4]. The distribution of betatron oscillation amplitudes remains, but it is no longer tied to the distribution of spin tunes. The spin tune distribution becomes more narrow and centered about an average value. Thus the requirements for an exceptionally long lifetime for the IPP become beam bunching, electron cooling, and sextupole field adjustment.

At the time that these data were recorded, other measurements were made under similar machine conditions but with the electron cooling left running continuously during the machine cycle. These data have recently been analyzed with the goal of quantifying any effect that the additional electron cooling has on the IPP lifetime. The analysis is described in the next section; afterward all the results are presented for comparison.

These data were not included in Ref. [3] since electron cooling is thought to be incompatible with the EDM search, both because of the solenoidal magnetic field needed for electron beam confinement and possibly uncontrollable systematic errors caused by any electron beam current induced precession. In other storage ring experiments with electron cooling, extra solenoids are often introduced to cancel the main solenoidal field. But any residual solenoidal field generates a change in the vertical polarization component, mimicking an EDM signal.

## II. Analysis of additional IPP data

The data of Fig. 1 were part of a larger set of measurements made under optimal sextupole field conditions. This set of runs is listed in Table 1. In some cases electron cooling was used for a limited time prior to any spin manipulation (pre-cooled). In others, the electron cooling was operated for the entire beam store (fully cooled). As progress was made toward managing long beam storage times, the storing time in the runs was changed as noted by the cycle length. Out of this data, a total of seven runs were chosen for consideration, as shown in Table 1. The Fig. 1 data came from run 5126.

Table 1: Long IPP lifetime runs included in analysis

| RUN NUMBER | E-COOLING TYPE | CYCLE LENGTH (s) | IPP LIFETIME (s) | TOTAL ERROR |
|---|---|---|---|---|
| 5018 | Fully cooled | 564 | 1721 | 1044 |
| 5019 | Fully cooled | 564 | 1263 | 325 |
| 5021 | Pre-cooled | 564 | 436 | 68 |
| 5023 | Fully cooled | 1562 | 2108 | 714 |
| 5039 | Fully cooled | 1564 | 2234 | 523 |
| 5126 | Pre-cooled | 1564 | 825 | 108 |
| 5127 | Fully cooled | 1564 | 987 | 180 |

The analysis methods have been described elsewhere [3,4].

In normal operation, a steady stream of particles are continuously taken or extracted from the beam and brought to the front face of a carbon block target from which they may scatter into the polarimeter detectors [3,4]. Extraction was achieved by applying a vertical white noise electric field with a frequency spectrum concentrated near a harmonic of the vertical beam tune. This brings beam particles with particularly large betatron amplitudes to a carbon block target located above the beam. In order to preserve the beam over a long storage cycle, the extraction of the beam onto the polarimeter target was made in four sampling periods of 15 to 30 s duration. An example of the data acquisition rate is shown in Fig. 2. The zero on the time axis corresponds to start of data acquisition which is just before the rotation of spin into horizontal plane.

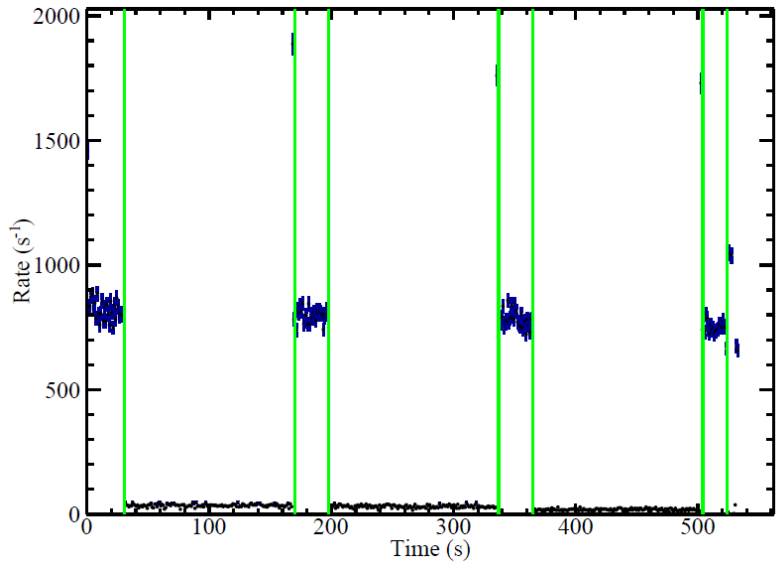

Figure 2: Example of the up detector rate in a single machine cycle as a function of the time in the cycle. The origin of the horizontal axis corresponds to the beginning of the first extraction period; the initial part of the cycle devoted to beam preparation is not shown. The measurements are from run 5021. The vertical lines mark the limits of the data chosen for analysis. A spike in the rate leading to a high rate point always occurs whenever the white noise extraction is first turned on.

The data taken with the polarization in the horizontal plane were divided into one-second time bins. Within each time bin, the direction of the IPP for each event could be assigned based on an assumed value for the spin tune and a time-stamp indicating when the event was recorded [9]. Then the events were sorted into 9 groups based on their polarization direction, which is given by the total rotation angle of the polarization since the start of data acquisition modulo $2\pi$. The range from 0 to $2\pi$ is then divided into 9 shorter intervals or groups covering all directions around a circle in the horizontal plane. Within each group the value of the up-down asymmetry was calculated for the one-second collection of events. As a function of direction, this produces a sinusoidal up-down asymmetry, as shown in Fig. 3 for two different values of the IPP. It is assumed that the actual spin tune value and the value used in the analysis match. The IPP is the magnitude of this sinusoid.

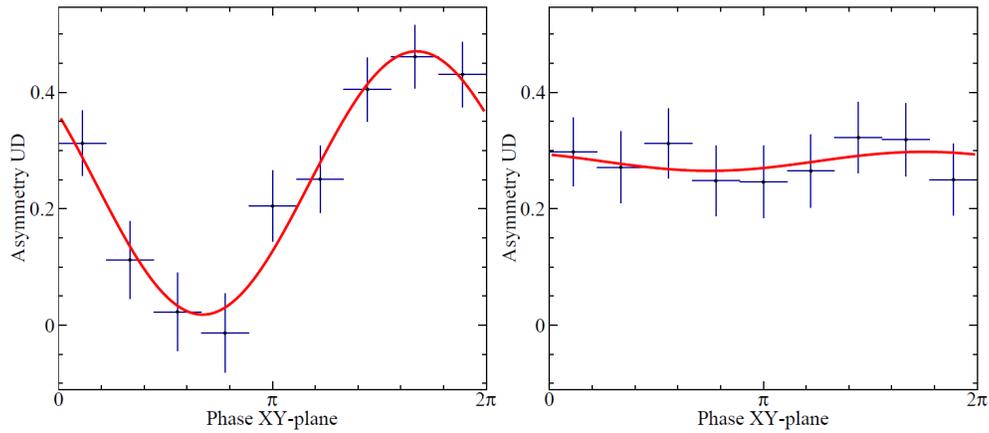

Figure 3: Measurements of the up-down polarimeter asymmetry for a set of 9 directional bins during a 1-second time interval with horizontal polarization in the COSY ring. The two panels contain examples of a large and a small IPP. The red curve shows the best fit of a sinusoidal curve through these data. Its magnitude is proportional to the size of the IPP. The offset of the sine curve from zero arises from a geometrical misalignment of the polarimeter relative to the beam direction and does not affect the magnitude of the sine curve.

    This procedure only works if the spin tune value used in the analysis is chosen with enough precision to maintain the integrity of the event direction sorting across the 1-second time interval. In practice the value of the spin tune is not known at first with sufficient precision. In the analysis a series of spin tune values lying in a narrow range about the expected spin tune were tried. The best choice is found by locating the spin tune which yields the largest value of the IPP magnitude $M$ in the formula [9] $M \sin(2\pi\nu_S + \varphi_S) + \rho$. In addition to the magnitude, the fit also produces a phase ($\varphi_S$) for the sinusoid relative to the start time for the store and the systematic offset $\rho$.

    Figure 4 shows two examples of the phase data from different beam stores. Each point represents the phase obtained from the analysis illustrated in Fig. 3 for a 1-second time interval. Data between the extraction periods shown in Fig. 2 are not shown. By convention, all of the points lie between $-\pi$ and $\pi$; as points get close to a limit they may be plotted close to the opposite limit when in fact they are really part of a smooth trend (see top panel).

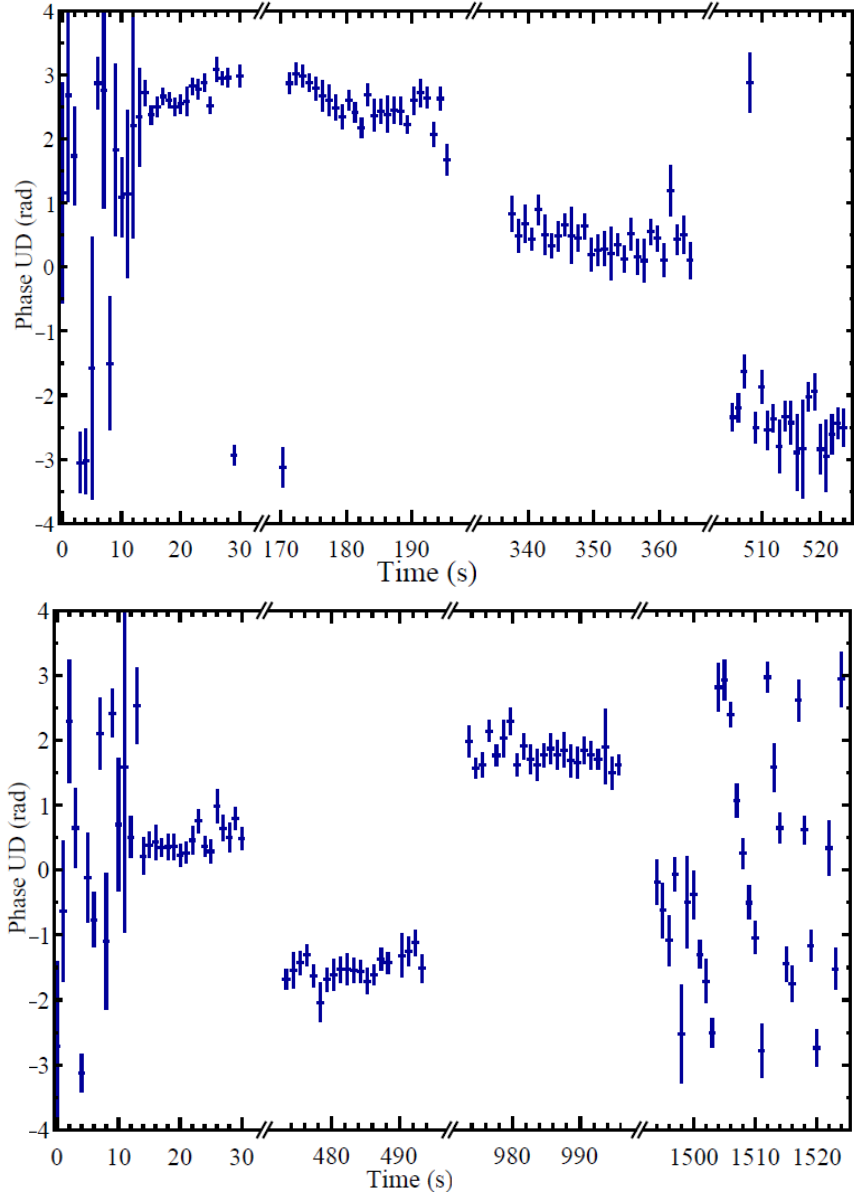

Figure 4: (upper panel) Example of up-down phase data for one cycle in run 5021. (lower panel) Example of the up-down phase data from run 5039 for a cycle that was discarded from the analysis. Parts of the time axis without usable data have been removed. See text for more details.

The phase is erratic at the beginning before the polarization has been rotated into the horizontal plane, which happens at about 13 s relative to the data acquisition starting time. After that, values of the phase are generally well determined if there are sufficient statistics and the values follow a smooth trend. In the top panel, the IPP phase remains relatively stable during each extraction period. In the lower panel the last extraction period shows a rapidly moving phase. This means that, for some reason, the spin tune is not sufficiently close to constant in this beam store. This complicates the assignment of a polarization magnitude. Stores such as this were discarded from the final data analysis.

The erratic phase at the beginning of the panels in Fig. 4 is an example of what happens when the IPP is close to zero. The phase will take on whatever value is needed to maximize the IPP, and the IPP result is thereby too large [9,10]. In the analysis, this systematic overestimate was suppressed by fixing the phase in advance to a local average of nearby phases. But the IPPs determined during each of the four extraction periods used here were generally too large to be significantly affected by this problem. So there is no need to make a correction for this systematic error.

The results for individual runs will be discussed in the next section.

### III. Results for remaining runs with long IPP lifetimes

The measurements for all of the runs are shown in Figs. 5 and 6. The results for the IPP time curves have been divided into two groups. Figure 5 contains the two runs where there was only electron pre-cooling. In these cases, the cooling was off while data was being accumulated. Figure 6 contains all of the cases where the cooling ran continuously through the data taking time. The IPP lifetime is taken from a template curve scaled horizontally and vertically to match the measurements. In some cases, the freedom given to the scaling along the vertical axis makes the initial value of the template curve differ noticeably from one. But in doing so, it allows this degree of freedom to be used to assess the errors in the IPP lifetime, as described below. All of the plots in Figs. 5 and 6 use consistent polarization and time axes to facilitate comparison.

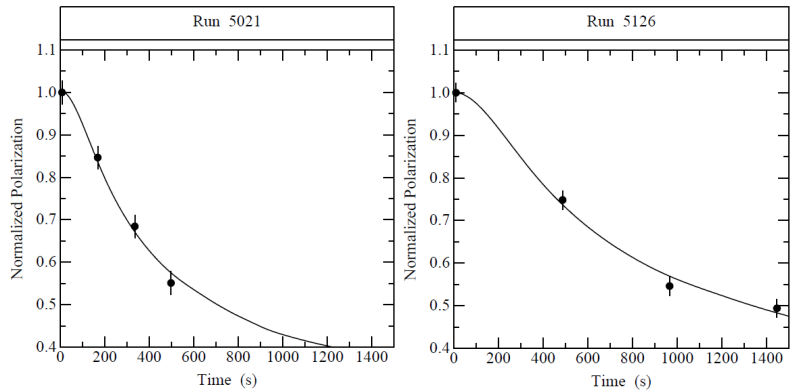

Figure 5: Measurements of the normalized IPP as a function of time for runs 5021 and 5126. Both runs had only electron pre-cooling.

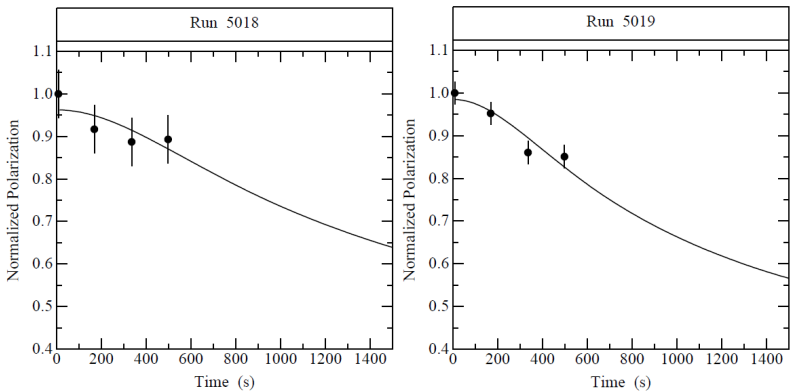

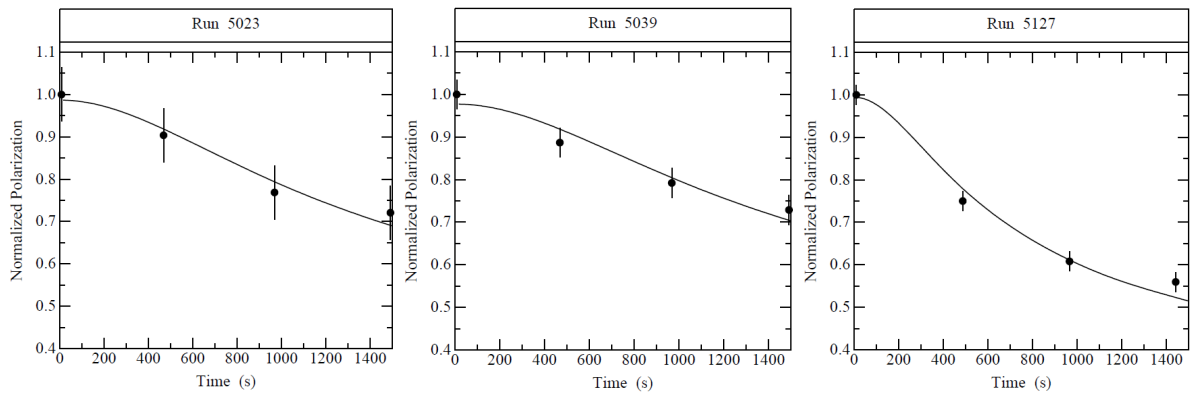

Figure 6: Measurements of the IPP as a function of time. The top row shows runs 5018 and 5019, both of which had the shorter (564 s) data observation length. The second row shows measurements of the IPP for runs 5023, 5039, and 5127 where the longer observation time was applied. In all five cases, electron cooling was operating throughout the data acquisition period.

The IPP clearly falls faster when there is only pre-cooling. Both of these measurements result in a smaller lifetime (see Table 1) than any case for continuous electron cooling. This effect is consistent between the runs with short (564 s) and long (about 1564 s) data taking times.

In the original paper containing Fig. 1, the IPP half-life was quoted along with the lifetime obtained as a "Gaussian" width (where the IPP falls to 60.6% of its original value). In Table I, all of the lifetimes are based on the Gaussian width definition. For the comparison here, a lifetime is less well determined from the template curve if there are no data points that straddle the 60.6% IPP value. This is the case for runs 5018, 5019, 5023, and 5039.

A single template is used for the polarization lifetime curve in each panel of Figs. 5 and 6. Adjustments are made mainly to the scale for the horizontal axis in order to obtain a best representation of the data for each case. The shape of the curve represents the evolution of a distribution of spin tunes that include contributions from unfolding a profile of the circulating COSY beam and an additional group centered near the reference spin tune representing particles whose orbits were corrected by the COSY sextupole fields. The features of this shape include a rapid initial fall followed by a flattening of the curve at larger times. These features appear significant only for runs 5126 and 5127, but there is not enough sensitivity to the two model components to determine their relative size with any precision. Thus a mixture that yielded a good approximation for runs 5126 and 5127 was chosen and used for all cases. The only important feature of this shape is that it fits the data well enough to allow an estimate of the IPP for each run.

The IPP shape was stored as a table of 2000 values. The adjustment to reproduce the data was made by scaling the shape vertically and horizontally while keeping the starting point of the shape fixed at the origin of the graph. The two scaling factors were adjusted until the chi square of the deviation between the curve and the IPP data was minimized. This procedure generated an error estimate for each of the two scaling factors. The error in horizontal scaling, which was proportional to time, was calculated for the point where the IPP curve became equal to 0.606. An additional error that was generated for this point by adjusting the vertical scaling by its fitting error was also included as a second contribution. Of the two error contributions the former is dominant by a factor 3 to 5.  Table 1 contains the total error which is their sum in quadrature.

The shape of the template curve chosen has a smooth fall away from one, followed by a leveling of the polarization value below about 0.5. This shape tends to work well for runs such as 5126 and 5127 where the data also contain these two features in the shape. For the remaining cases, this shape suffices even though both features are not present in the measurements. A straight line would do as well in most instances. This is particularly true for those cases where the polarization values did not fall below 0.7. Thus these measurements have limited information on the time dependence of the polarization loss.

For runs 5126 and 5127, the flattening of the shape at larger times would suggest that there is an outer part of the beam (since this part is extracted first at the carbon target [1]) whose IPP lifetime is limited while the inner core of the beam (extracted later) has a much longer IPP lifetime.

The clear effect of continuous electron cooling on the IPP lifetime suggests that this would be an important feature of any ring that might be used for an EDM search. However, the electron beam is usually confined during cooling by a solenoidal field that runs the length of the cooling region. This would create a precession of the beam polarization that would destroy the sensitivity to an EDM. Thus electron cooling must be avoided. It remains to be tested whether or not stochastic cooling avoids this difficulty.

## IV. Conclusions

When the polarization direction of a stored, polarized beam is placed in the ring plane, variations in the betatron oscillation amplitudes and consequently the speeds of the particles under bunched beam conditions lead to decoherence causing an eventual loss of polarization. In part, this loss may be controlled by reducing the phase space occupied by the beam through electron cooling. The IPP lifetime may be further improved by applying sextupole field corrections to the ring, particularly in the arcs. These corrections deform the orbits of particles whose betatron oscillation amplitudes are large in such a way as to reduce average orbit radius and hence the circumference, thus further extending the IPP lifetimes to times longer than 1000 s.

A survey of additional data that happened to include tests with electron cooling operational throughout the beam store showed a clear improvement in the IPP lifetime, which suggests that cooling should be an important feature of any ring that might be used for experiments relying on a stable IPP with a long lifetime, such as an EDM search or a proposed axion hunt [11,12]. Systematic issues with electron cooling in an EDM search raise the question of whether other cooling mechanisms should be investigated.

**Acknowledgements:**

The authors wish to thank the members of the JEDI collaboration for their help with this experiment. We also wish to acknowledge the staff of COSY for providing good working conditions and for their support of the technical aspects of this experiment. This work has been financially supported by by an ERC Advanced-Grant (srEDM #694340: "Electric Dipole Moments using storage rings") of the European Union, and by the Shota Rustaveli National Science Foundation of the Republic of Georgia (SRNSFG grant No. 217854: "A first-ever measurement of the EDM of the deuteron at COSY"). Participation of S.K. in the project was supported within the DSC 2019 grant for young researchers and Ph.D. students of the Faculty of Physics, Astronomy and Applied Computer Science of the Jagiellonian University, the Polish Ministry of Science and Higher Education No: 2019-N17/MNS/000002. The work of N.N. was part of the Russian Ministry of Science program 0033-2019-0005.